\documentclass[fleqn,usenatbib]{mnras}
\usepackage{newtxtext,newtxmath}

\usepackage[T1]{fontenc}
\usepackage{ae,aecompl}

\usepackage{graphicx}	
\usepackage{amsmath}	
\usepackage{natbib}
\usepackage{graphicx}
\usepackage{caption}
\usepackage{subcaption}
\title[Bulge to total luminosity ratio using deep learning]{Predicting bulge to total luminosity ratio of galaxies using deep learning}

\author[Grover et al.]{
Harsh Grover$^{1}$\thanks{E-mail: grover.harsh.bitsp@gmail.com (HG)},
Omkar Bait$^{2}$, Yogesh Wadadekar$^{2}$ and Preetish K. Mishra$^{2,3}$ 
\\
$^{1}$Birla Institute of Technology and Science, Pilani, Rajasthan 333031, India\\
$^{2}$National Centre for Radio Astrophysics, TIFR, Post Bag 3, Ganeshkhind, Pune 411007, India\\
$^{3}$Inter-University Centre for Astronomy and Astrophysics, Post Bag 4, Ganeshkhind, Pune 411007, India
}

\date{Accepted XXX. Received YYY; in original form ZZZ}

\pubyear{2019}

\begin{document}
\label{firstpage}
\pagerange{\pageref{firstpage}--\pageref{lastpage}}
\maketitle

\begin{abstract}

We present a deep learning model to predict the $r$-band bulge-to-total light ratio ($B/T$) of nearby galaxies using their multi-band JPEG images alone. Our Convolutional Neural Network (CNN) based regression model is trained on a large sample of galaxies with reliable decomposition into the bulge and disk components. The existing approaches to estimate the $B/T$ use galaxy light-profile modelling to find the best fit. This method is computationally expensive, prohibitively so for large samples of galaxies, and requires a significant amount of human intervention. Machine learning models have the potential to overcome these shortcomings. In our CNN model, for a test set of 20000 galaxies, 85.7 per cent of the predicted $B/T$ values have absolute error (AE) less than 0.1. We see further improvement to 87.5 per cent if, while testing, we only consider brighter galaxies (with $r$-band apparent magnitude < 17) with no bright neighbours. Our model estimates $B/T$ for the 20000 test galaxies in less than a minute. This is a significant improvement in inference time from the conventional fitting pipelines, which manage around 2-3 estimates per minute. Thus, the proposed machine learning approach could potentially save a tremendous amount of time, effort and computational resources while predicting $B/T$ reliably, particularly in the era of next-generation sky surveys such as the Legacy Survey of Space and Time (LSST) and the  Euclid sky survey which will produce  extremely large samples of galaxies.

\end{abstract}

\begin{keywords}
galaxies: structure -- methods: data analysis -- techniques: image processing

\end{keywords}



\section{Introduction}


Artificial Intelligence (AI) is the domain of computer science that deals with computers performing tasks which usually need human-level intelligence. Machine Learning (ML) is the branch of AI in which programs learn from experience to progressively get better at a task \citep{Mitchell97}. A ML model develops its own implicit set of rules for a certain task using a learning algorithm with experience. When learning from labelled datasets, a branch of ML called supervised learning is used. The aim of supervised learning is to learn a latent function that maps the inputs to the ground-truth labels in a given dataset. Depending on the distribution of ground-truth labels, supervised learning is broadly divided into classification and regression problems. If labels belong to a discrete space (or are categorical) the problem is defined as a classification problem. Whereas, if they belong to a continuous space, the problem is defined as a regression problem.

Over the years, many studies have utilised the competence of ML for classification, parameter estimation and outlier detection problems in data-driven astronomy \citep[see][for a detailed overview]{Baron19}. One of the popular supervised learning algorithm is Support Vector Machines (SVM) which has been applied successfully to extract useful information from the data coming from large sky surveys. Particularly in the field of extragalactic astronomy, ML and SVM based techniques have been used for star-galaxy classification \citep{Philip2002}, photometric redshifts \citep{Wadadekar_2005}, galaxy morphology classifications \citep{refId0} etc. In recent years, a new ML technique -- Deep Learning -- has emerged as the best performing supervised learning technique across many applications in different disciplines \citep{dl2015}.

Deep learning (DL) is the paradigm of machine learning which uses multi-layer neural networks. Neural networks are ML models inspired from the network of brain cells or neurons. The differentiating factor between DL and conventional ML is in the process of feature selection. In conventional ML, performance strongly depends on the features used. More often than not, new features are created for a task to better capture correlations in the data. This is called feature engineering. ML models perform poorly if not given good features to learn from. Deep learning eliminates this dependence by defining its own features to learn from, which are relevant to the task at hand. This makes deep learning a versatile, high performing solution for a variety of supervised ML tasks. The Artificial Neural Networks (ANNs) have found application in not only the traditional tasks in extragalactic astronomy such as photometric redshift estimation \citep[e.g.,][]{Firth03, Collister04, Tagliaferri03, Vanzella04, Sadeh16, Bilicki18, Pasquet19}, but also in more specialised problems such as predicting infrared luminosity of galaxies \citep{Elison2016}, estimation of star formation properties \citep{2020MNRAS.493.4808S} and ranking the quenching parameters of galaxies \citep{Teimoorina2016}. In particular, Convolutional Neural Networks (CNN) are now extremely popular in  studies using imaging data on galaxies. The CNNs are deep learning models designed to extract features from images. They have provided state-of-the-art performance in majority of computer vision tasks in recent times \citep{alexnet2012, he2017mask}. CNNs have been utilised for galaxy morphological classification e.g., classifying the optical morphologies broadly into spheroidal, disk and irregular types \citep{Huertas-Company15} and even for classification of various radio galaxy morphologies \citep{2019MNRAS.482.1211W}. Recently, \citet{Ribli19} have shown that CNNs can be used to predict galaxy shapes, needed for weak lensing studies, from a wide field but shallow sky survey images using the 'ground truth' images from a deeper but narrower field survey. CNNs have also found their application in the automated detection of features in sky-survey images such as galactic bars \citep{Abraham2018} and strong gravitational lenses \citep{Li2020, Canameras2020}. Further, ML and CNNs in particular have also been used to detect outliers in large area Sloan Digital Sky Survey (SDSS) data \citep[e.g.,][]{Baron17, Sharma19}. In this study, we want to determine the bulge to total luminosity ratio ($B/T$) of a galaxy using its optical multi-band images as input. Due to the use of a dataset with galaxy images labelled by $B/T$ $\in [0,1]$ in a continuous space $\rm I\!R$, we have performed CNN based regression in this work. 

The bulge to total light ratio ($B/T$) of a galaxy is the fraction of total flux coming from the bulge component of the galaxy. Bulges are the central component of disc galaxies which appear as central bright cores in galaxy images or as excess-light over the disc light in the inner region of galaxy light profiles. The $B/T$ ratio is an important indicator of galaxy structure and is well correlated with several other physical quantities of interest in studies of galaxy evolution, such as galaxy morphology \citep{Graham2008}, kinematics \citep{Cappellari2013}, stellar mass and star formation rate \citep{Bluck2014}. The $B/T$ is also directly related to the bulge luminosity which in turn correlates with the mass of the central supermassive black hole of the galaxy \citep{Kormendy1995, Marconi2003,Kormendy-Ho2013}. Galaxies having different $B/T$ ratios are thought to have undergone evolution along different evolutionary paths. A high value of $B/T$ of a galaxy generally indicates its evolutionary history dominated by galaxy mergers \citep{Hopkins2015}. On the other majority of low $B/T$ systems are often pseudobulge hosting galaxies \citep{Fisher2008, Gadotti2009} which are thought to undergo slow evolution through internal, secular evolution processes \citep{Kormendy2004}. It must be noted that unlike visual morphology which is a qualitative measure, $B/T$ is a quantitative measure of the galaxy morphology. It is the only reliable way to separate ellipticals and disc galaxies when  morphological features such as spiral arms cannot be resolved by the telescope (due to resolution or sensitivity limitations, particularly at high redshifts). The quantitative nature of the $B/T$ ratio also allows direct comparison of galaxies with their counterparts in cosmological simulations. One can study these simulated galaxies to understand the origin of various properties of real galaxies. All these factors make the $B/T$ an important parameter to measure in studies of galaxy formation and evolution. 

The $B/T$ ratio of galaxies has been traditionally measured by modelling their surface brightness profile. Such modelling allows one to find the light-profiles of individual photometric components of galaxies and then measure their properties such as luminosity, shape, size etc. The earliest works aimed at photometric decomposition of galaxies involved fitting for the azimuthally averaged 1-D surface brightness profile \citep{Kent1985}. However, this method leads to systematic errors as creation of 1-D light profile by either averaging over the galaxy image or by taking a single cut along the major axis do not properly take into account non-axisymmetric features like a bar or isophotal twist (for more details see \cite{Gao2017}). Due to these shortcomings, techniques to fit the whole 2D image of the galaxy, pixel-by-pixel, were developed \citep{Wadadekar1999}. There are now a number of implementations of 2D bulge disk decomposition available  -- \textsc{GIM2D}: \citet{GIM2D}, \textsc{GALFIT}: \citet{GALFIT}, \textsc{IMFIT}: \citet{IMFIT}, \textsc{ProFit}: \citet{ProFit}, \textsc{BUDDA}: \citet{BUDDA}, \textsc{GASP2D}: \citet{GASP2D} --  which fit the 2D images of galaxies. They usually employ different algorithms and methods to find the best fit model for a galaxy image. In the last two decades, some pipeline codes have been developed (\textsc{PYMORPH}: \cite{PYMORPH}, \textsc{GALAPAGOS}: \cite{GALAPAGOS}) to carry out bulge disc decomposition for large galaxy samples, in a mostly automated fashion. This 2D bulge-disc decomposition approach, while being powerful, is time consuming. The employed fitting algorithm and code carries out a search in parameter space defined by various model parameters and then finds the best fit. Most of the time one also needs to carry out additional steps (masking and providing PSF images) before the actual fitting of galaxies take place. Thus estimating $B/T$ for a single galaxy involves a significant amount of time and computational cost. More importantly, the fitting procedure scales only linearly with the size of the galaxy sample. This limitation will be a major problem in the era of next generation sky surveys (e.g., LSST, Euclid sky surveys) which will observe several billion galaxies. 

For an accurate and fast estimation of $B/T$ for large galaxy samples, one can adopt a Machine Learning approach. One can develop a ML algorithm to learn the latent mapping from the galaxy images to the $B/T$ ratio which will reduce the inference time by orders of magnitude. The only time consuming part in this approach is training the model, which is not an issue because it has to be done only once. Such a ML based approach has been found to be useful in estimating galaxy structural parameters in the past. For example, \cite{Tuccillo_2017} have used CNN to fit the surface brightness profile of galaxies.  This study demonstrates the computational efficiency and speed of deep learning over conventional fitting methods. However, their analysis is limited to single component (S\'ersic) fit. In this study, we propose a CNN based regression model to estimate $r$-band $B/T$ ratio of galaxies using their $g$,$r$,$i$-band images from the SDSS. We have used images of galaxies coming from the SDSS legacy survey DR 15 \citep{SDSSDR15} and their $B/T$ values as listed in \cite{meert} which is a large bulge-disc decomposition catalogue of SDSS galaxies. 

The organisation of this paper is as follows. In Section \ref{data} we describe the data used in our study. In Section \ref{cnn}, we describe the deep learning model and the training methodology used for the analysis. In Section \ref{rnd}, we discuss the results of our study with various metrics and error analysis. We end this paper by providing a summary and  conclusion in Section \ref{conclusion}.

\begin{figure*}
     \centering
     \includegraphics[scale=1]{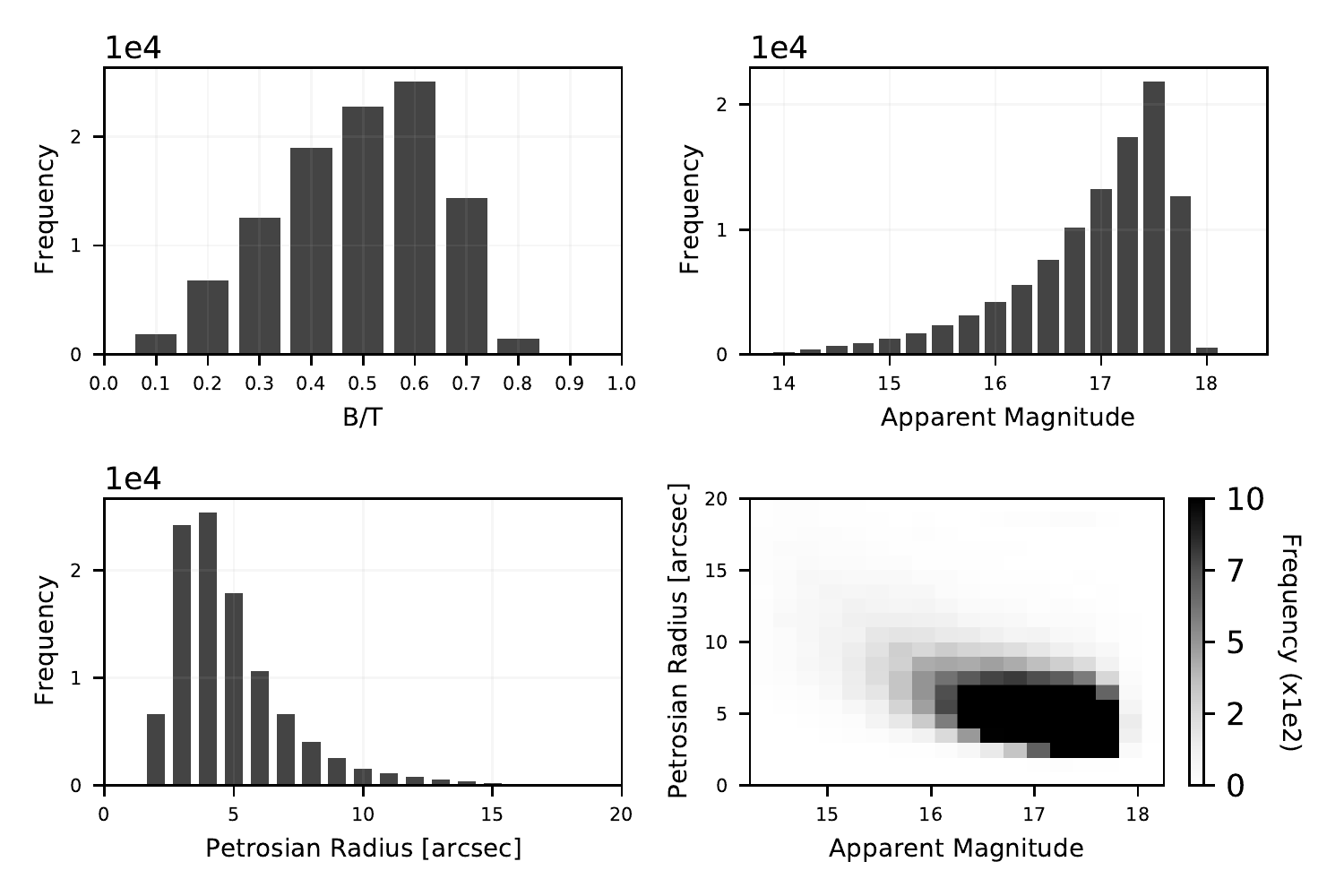}
     \caption{Distribution of $B/T$, apparent magnitude and Petrosian radius for our galaxy sample as measured in the SDSS $r$ band. We selected galaxies based on good two component fits as described in Section 2.1. Our sample of \textasciitilde103,000 galaxies is about 15 per cent of the Meert catalogue.}
     \label{fig:dist}
 \end{figure*}

\section{Data and sample selection} \label{data}

In order to train and test our CNN based model capable of predicting $B/T$ ratio of galaxies, we needed a statistically significant number of galaxies with available $B/T$ measurements. To select our sample for this purpose, we started with the data from \cite{meert} (hereafter Meert catalogue) which is a catalogue of 2D photometric decompositions of about $\sim$ 670,000 spectroscopically selected galaxies coming from Sloan Digital Sky Survey (SDSS) Data Release 7 \citep{Abazajian09}. This is a flux limited sample of low-redshift galaxies lying within the redshift range $0.05<z<0.3$ and having $r$-band Petrosian magnitude between $14.0 < r < 17.77$. The fits in the Meert catalogue are performed on $r$-band SDSS images using the \textsc{GALFIT} code with the \textsc{PYMORPH} analysis pipeline \citep{PYMORPH}. Each galaxy in this catalogue is fitted by four different models of galaxy light profiles, two of which are single component de Vaucouleurs and S\'ersic profiles and the other two are two-component de Vaucouleurs+Exponential and S\'ersic+Exponential profiles. The galaxies having bad fits, or having erroneous magnitude and size measurements are indicated by setting bit 14 and 20 in the 'finalflag' column in the Meert catalogue. The galaxies having good fits in this catalogue are divided into three categories: those having good bulge fits only, the ones with good bulge+disc fits and those having good disc fits only. Since the aim of this work is to predict the $B/T$ ratio of galaxies, we chose to work with galaxies having good two component bulge+disc fits only. These galaxies have bits 10 AND 12 set in the 'finalflag' column of Meert catalogue. Additionally, we also removed galaxies having bulge S\'ersic index $\geq 8$ as recommended by \cite{meert} in order to exclude galaxies with unreliable measurements of subcomponent properties.  

Using the above mentioned constraints, we obtained a sample of 103,757 galaxies having measurements of $r$-band $B/T$ ratio from the Meert catalogue. We treat these $r$-band $B/T$ ratio measurements as ground truth labels for our analysis.
As mentioned previously, we aim to come up with a CNN based regression model to predict the $B/T$ of galaxies via their $g$,$r$,$i$-band images. In order to complete the training and testing requirement of our model, we downloaded a $gri$ colour composite image in JPEG format using the SDSS finding chart tool \footnote{https://skyserver.sdss.org/dr15/en/tools/chart/chartinfo.aspx} for every galaxy in our sample. The size of each image is 128$\times$128 pixels. Scale of the image was kept at the natural scale of SDSS, which is 0.396 arcsec pixel$^{-1}$. Each image is therefore approximately, 50.7$\times$50.7 arcsec$^2$ in angular extent. All the subsequent analysis presented in this paper was done using these composite $gri$ images of 103,757 galaxies labelled by their respective $r$-band $B/T$ values. 

We tried both SDSS DR15 and DR7 images and found that our CNN performs better with DR15 images. Hence, all the  analysis in this paper uses DR15 images. This may be due to the fact that the JPEG images in DR15 are made using the Lupton colour scheme which is a more correct representation of galaxy colour \citep{Lupton04}. 

We also experimented with rescaling images to $128\times 128$ pixels in proportion to the Petrosian radius of each galaxy. However, our CNN performs better when natural scale is used for all images. We therefore, have used the natural scale images in our analysis.

We show some of the characteristics of our sample galaxies in Figure \ref{fig:dist}. The top-left panel of this figure contains the distribution of the $B/T$ ratio of our sample galaxies, which is seen to be fairly broad and peaking around $B/T=0.6$. Apparent magnitudes of galaxies in our sample range from 13.13 at the bright end to 19.16 at the faint end, however, most of them lie in the range of 16 to 18 as seen in the top-right panel of Figure \ref{fig:dist}. The bottom-left panel of Figure \ref{fig:dist} depicts the distribution of the Petrosian radius of galaxies in our sample. The full range of Petrosian radius goes from 1.61 arcsec to 68.31 arcsec but majority of them are smaller than 8 arcsec. Lastly, it should be noted that our requirement of a good two component fit excludes the extreme cases, pure ellipticals ($B/T=1$) and bulgeless disc galaxies ($B/T=0$) from our sample. Therefore, any result presented in this work should be taken to be valid for regular disc galaxies only.

\section{Convolutional Neural Networks} \label{cnn}

CNNs have now established themselves as the go to learning algorithm in computer vision, with state of the art performance in almost every task in this field  \citep{alexnet2012, he2017mask}. Since input to our model will be images, our task comes within the field of computer vision.

A CNN is a neural network with convolution layers. A convolution layer performs 2D convolution operations on the input image. The filters for these convolution are the learning parameter for the convolution layer. A typical CNN architecture has first few layers as convolution layers, followed by few fully connected layers \citep{lecun98}. As convolution operations work well for extracting image features, the part of the network containing convolution layers is also called Feature Extractor. 

The difference in using a feature extractor of a CNN and extracting features using regular image processing is that the filters in a convolution layer are learned by the network, based on the task. The first few layers extract low-level features, like edges and color shifts. As we go deeper in the network, high-level abstract features are extracted using the features from previous layers. These high-level task-specific feature extraction is what makes CNNs quite effective for computer vision tasks. 

The learning is generally done using backpropagation and gradient descent algorithms. The fully connected layers, which follows the feature extractor, flatten the different features learned into a single vector. The design of the final fully connected layer and the loss function used for training depends on the functionality required and varies from task to task. This last layer can be a classifier or can perform regression on the extracted features.

Since we are learning from galaxy images to predict $B/T$ $\in (0,1)$, we have used a CNN model for image regression. \textsc{Keras} \citep{chollet2015keras} -- an open source neural network library written in \textsc{Python} -- is used for building and training the CNN model.

\subsection{Model Architecture} \label{architecture}
A number of different CNN architectures exist based on the number of convolution layers, kinds of convolution performed by them and their placement in the network. We tested the architectures with good performance on ImageNet ILSVRC challenge \citep{ILSVRC15}, i.e. InceptionV3 \citep{incv3}, Resnet50\citep{he2015deep}, Xception \citep{xception} and Inception-ResnetV2 \citep{DBLP:journals/corr/SzegedyIV16}. After extensive experimentation we found that the \textit{Xception} architecture provides the best performance for our dataset. All the models were trained from scratch with random initialisation. Inception-ResNetV2 performed slightly worse than Xception but with more than double the parameters. Therefore, Xception architecture was chosen due to its high performance as well as parameter efficiency.

\begin{figure*}
    \centering
    \includegraphics[scale=0.5]{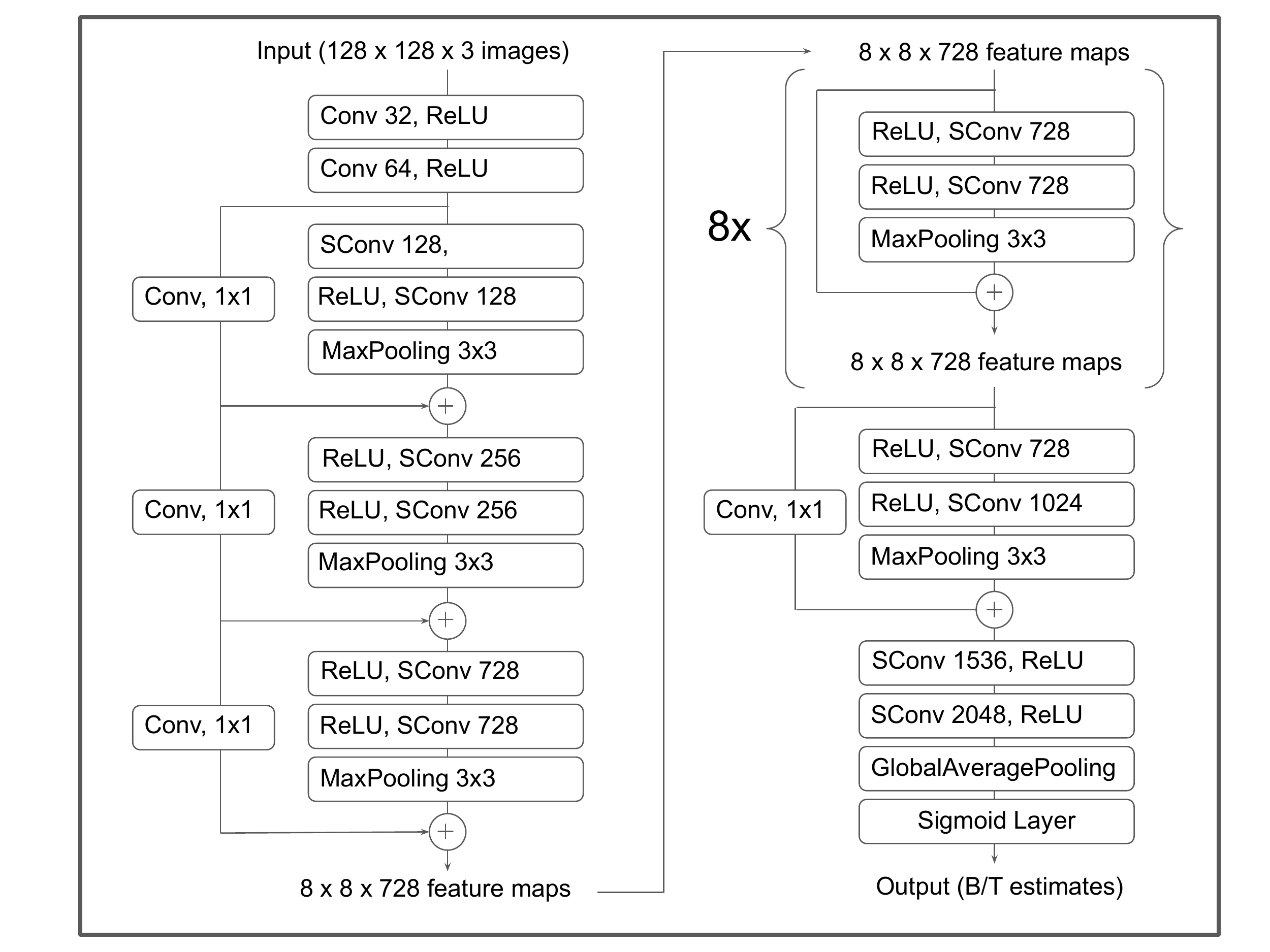}
    \caption[Xception architecture.]{Xception architecture \citep{xception} used in our work. SConv represents separable convolution. Number of filters are specified with every Conv and SConv operations. Every Conv and SConv operation uses filter size of 3x3 unless explicitly specified. Every Conv 1x1 and first Conv 32 use a stride of 2x2, rest all Conv and SConv use stride of 1x1.}
    \label{fig:xception}
\end{figure*}

The Xception network is a stack of depth-wise separable convolution layers with residual connections \citep{he2015deep}, as shown in Figure \ref{fig:xception}. Depth-wise separable convolution separates the spatial and cross-channel correlations completely, by breaking normal convolution operation into two sub-operations, point-wise and depth-wise convolution. Using two operations, instead of one normal convolution also results in fewer parameters and faster computation. Reduced number of parameters, in turn, reduces variance (overfitting), which is crucial for models with millions of parameters. \textit{Global Average Pooling} (GAP) as described in \cite{netinnet} is used after the Xception backbone to further reduce parameter count and variance without much loss of relevant information. The output of this layer is directly connected to a single output neuron with sigmoid activation. Sigmoid activation is used for the output layer because all possible values of $B/T$ $\in [0,1]$.

\subsection{Training Methodology}
\subsubsection{Data Pre-processing}
We processed the images before passing them as input to the model. This can be a crucial step to improve performance with the same model and dataset. We normalise the image pixel values for faster optimization during training. Data augmentation is also performed while training to regularise the model. A train-val-test split of 60-20-20 is used to quantify the training results and measure training performance after each epoch.

Data augmentation is a process to create more data samples for training from the existing samples. We do this by exploiting the invariances in our dataset. Galaxy images have rotational invariance, i.e. rotating an image by an arbitrary angle would not affect the B/T. Similarly, they are invariant to reflection, i.e. flipping the image horizontally or vertically would result in same B/T. We use these invariances to create multiple images from a single image. While training, every input image is rotated by a randomly selected angle between 0 and 360 degrees. They are also flipped horizontally and vertically with probability of 0.5 for each flip. We also normalise every pixel value to be in range [0,1], by dividing each value by 255, the maximal value of an 8 bit pixel. No other processing was done to the images. It is to be noted that in our dataset all the target galaxies are at the centre of the image. We used the  \textit{ImageDataGenerator} class in \textsc{Keras} to implement the augmentation and normalisation operations.

We use 20000 images each for test and validation sets. These images are chosen randomly from the 103757 galaxies selected as per section \ref{data}. The remaining 63757 samples constitute the training set. None of the images in the test or validation set are used for training. After each epoch, we validate the model on the validation set to measure its performance on unseen data. This also helps us to determine bias and variance of the model. This is necessary as we do not want the model to suffer from high bias (under-fitting) or high variance (over-fitting). We selected the model with best performance on the validation set, after the training. We then tested the selected model on the test set to gauge its performance in real world scenarios.

\subsubsection{Transfer Learning}

Pre-trained networks are known to converge much faster than a randomly initialised network with potential gain in performance for similar fine-tuning and pre-training datasets. In our case, because the pre-training dataset (imagenet) and fine-tuning dataset (galaxies) are not similar, we do not observe better performance by using pre-training. Hence, we have trained the model from scratch using random initialisation as our main model. However, one can still benefit from better weight initialisation from pre-training, which leads to faster convergence when training on our dataset. This in turn leads to much less use of computational resources to achieve comparable results. This trade-off is discussed further in Section \ref{performance}.

\subsubsection{Early Stopping}\label{es}
Early stopping is used to cease training after considerable over-fitting and to restore weights from the model with the best performance on the validation set. The number of epochs the system waits till there is no further improvement while training is called \textit{patience}.
 
\subsubsection{Tuning Hyperparameters}
\begin{enumerate}
    \item \textbf{Loss Function}: Loss function is a hyperparameter which determines how we want to train a model \citep{LeCun1988ATF}. It is a function of predicted values $\hat{\textbf{\textit{y}}}$ and ground-truth \textbf{\textit{y}}, and measures the difference between them. Mean Absolute Error (MAE) is a commonly used loss function for regression tasks. It was found to perform better than Mean Squared Error and Huber Loss, which are two other widely used loss functions used for regression. Therefore, we have used the MAE loss function.
    \item \textbf{Optimizer}: The aim of an optimizer is to find the global minimum for loss function in the parameter space. Adaptive Moment Estimation or Adam \citep{kingma2014adam} is currently the most widely used variant of Gradient Descent (GD) optimizer. It uses exponentially weighted average of gradients, or momentum, and root mean square propagation (RMSProp) with GD for adaptive estimation of lower-order moments. We have used Adam for our analysis. Learning rate of 5e-4 is used, which is determined empirically from the learning curves.
    \item \textbf{Mini-batch size}: Using the entire training set for one update of the optimizer can be impractical due to limited computational resources. It is more practical and faster to use only a subset of training examples for GD update. This is called Mini-batch Gradient Descent and the subsets of training set are called mini-batches. We determine the mini-batch size as the largest number of samples which can be processed without exhausting our GPU RAM. We have used 128 images as the mini-batch size due to limited GPU RAM.
    \item \textbf{Early stopping patience}: As described in Section \ref{es}, patience determines number of epochs the system waits after there is no improvement while training. It is determined empirically from the learning curves, based on the learning rate. We never observed  the loss decrease after over-fitting for more than 15 epochs, without using early stopping. Therefore, we used a patience of 20 epochs.
\end{enumerate}



\section{Results and Analysis} \label{rnd}
\subsection{Performance Metric} \label{metric}
The metric used to determine the performance of the model is based on absolute error in prediction of the $B/T$ ratio. Performance is measured as the fraction of test samples with $AE$ less than a threshold absolute error ($AE$). We have defined accuracy as performance for $AE$ cut-off = 0.1. This choice of the $AE$ cutoff is somewhat arbitary; with threshold $AE=0.1$,  for a galaxy with true $B/T = 0.4$, we label the prediction a success if the predicted value $(B/T)_P$ is $0.3 < (B/T)_P < 0.5$

\subsection{Learning Curve}
\begin{figure}
    \centering
    \includegraphics[scale=0.5]{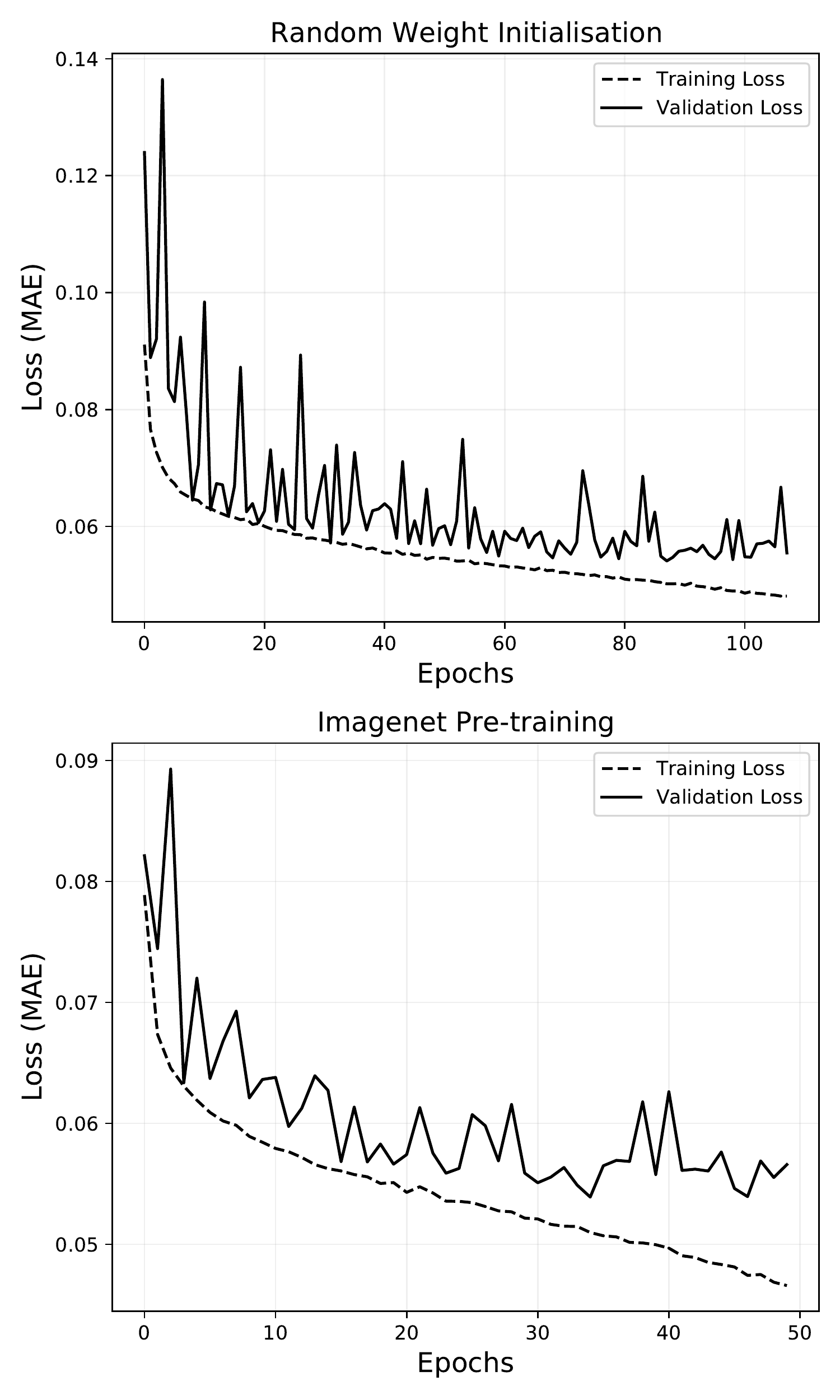}
    \caption{Learning Curves for the Xception model trained using random weight initialisation and Imagenet pre-training. The dashed curves show changes in MAE loss for the training set with each epoch. The solid curve shows the same for validation set. }
    \label{fig:lr_curve}
\end{figure}

The curves in Figure \ref{fig:lr_curve} show training and validation loss for each training epoch with learning rate = 5e-4. The first curve is for Xception model trained from scratch and the second curve is for pre-trained Xception model. We have trained using Google Colaboratory\footnote{https://colab.research.google.com/} environment with NVIDIA Tesla T4 GPU (12.7 GB GPU RAM) and 12 GB CPU RAM. Each epoch takes about seven minutes to complete with the specified hardware. Early stopping is used with patience of 20 epochs. We observed the lowest validation loss value of 0.054 at epoch 89 for the Xception model trained from scratch with random initialisation. After that the validation loss stopped decreasing. The learning stopped at epoch 109 with the use of early stopping, as the model had started over-fitting. The weights at epoch 89 are hereby used for analysis. The complete training took close to 13 hours.
The pre-trained model is observed to converge much quicker with the same learning rate. The pre-training was done on ImageNet ILSVRC challenge dataset \citep{ILSVRC15}. This subset of ImageNet dataset contains over a million labelled images from 1000 classes. We then retrain the whole model with our training data. We observed the lowest validation loss value of 0.0539 at epoch 39 for the pre-trained Xception model.

\subsection{Bulge-to-Total Luminosity Ratio}
\subsubsection{Performance} \label{performance}
\begin{figure}
    \centering
    \includegraphics[scale=0.5]{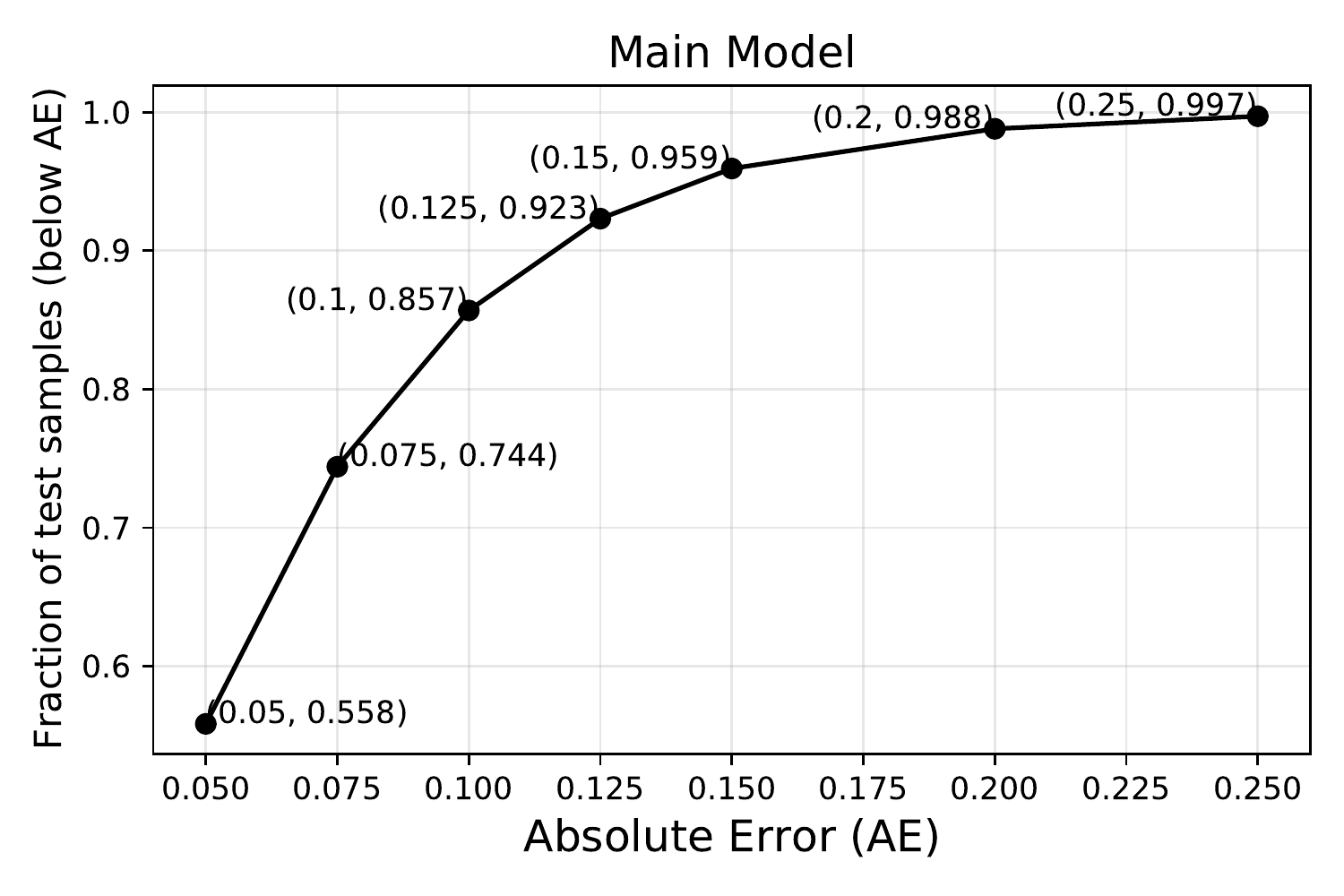}
    \caption[Performance as a function of absolute error cut-off.]{Performance as a function of absolute error cut-off. For any point $(x,y)$ on this graph, \textit{x} is the absolute error and \textit{y} represent fraction of test samples with absolute error less than x. On the entire test set of 20000 images, we achieved accuracy of 85.7 per cent with the main model. Accuracy is defined as the fraction of test samples with absolute error less than $0.1$.}
    \label{fig:result}
\end{figure}

With the Xception model trained from scratch with random initialisation, we record an accuracy of 85.7 per cent and MAE of 0.0543 on the test set. (see fraction (y axis) at absolute error (AE) $= 0.1$ in Figure \ref{fig:result}). This model is subsequently referred as the main model. The accuracy with InceptionV3, Resnet50 and Inception-ResnetV2 are 84.7, 84.9 and 85 per cent respectively. These models are also trained from scratch with random initialisation.

Using a pre-trained Xception model, we record an accuracy of 85.2 per cent and MAE of 0.055 on the test set.  The pre-trained model reaches loss minimum in just 39 epochs as opposed to 89 epochs for the main model, as shown in Figure \ref{fig:lr_curve}. There is hence a trade-off between the performance increase and the additional computational resources required to achieve the higher performance. The test set is kept the same for both pre-trained accuracy measurement and  main model accuracy measurement.


The main model takes 44.1 seconds to predict $B/T$ for the 20000 test images on an NVIDIA Tesla P100 GPU with 12.7 GB GPU RAM. On a more modest Intel Xeon CPU at 2.3GHz with 12 GB RAM, the main model takes 17 minutes to predict $B/T$ ratio for the 20000 test images.

It is important to test the sensitivity of our predictions to the fact that the galaxy of interest is always located at the centre of the image. Shifting the target galaxies during testing, by as few as 10 pixels in either X or Y direction reduces the accuracy to $\sim$55 per cent and shifting them by 20 pixels reduces it further to $\sim$30 per cent. Note that with our acceptance criterion of $AE < 0.1$ and with $0 < B/T < 1$, about 20 per cent accuracy would be achieved by random chance, even without any learning. Thus the strong dependence of accuracy in $B/T$ ratio prediction on the distance of the target galaxy from the centre, shows that our model is not shift-invariant and has learned the position of the target galaxies. 

\begin{figure*}
    \centering
    \includegraphics{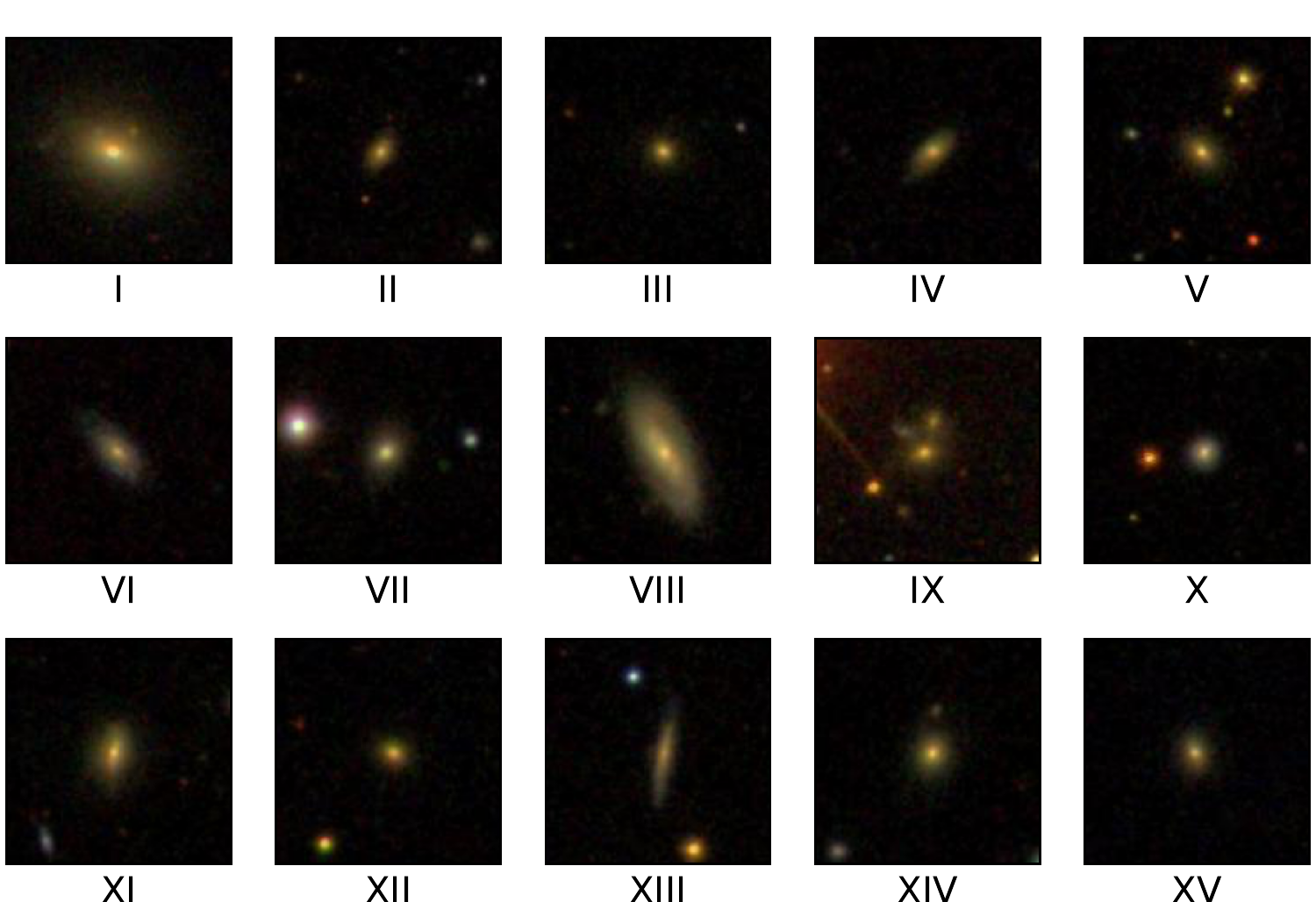}
    \caption{Predicted $B/T$ of 15 random galaxies from our test sample using the main model. Refer Table \ref{tab:res_main} for True $B/T$, Predicted $B/T$ and Absolute Error for these galaxies.}
    \label{fig:res_main}
\end{figure*}

Some randomly sampled galaxies with their real and predicted $B/T$ values are shown in Figure \ref{fig:res_main} and Table \ref{tab:res_main}. It is clear that the $B/T$ ratio is estimated well over a wide range of galaxy size, brightness and morphology.

\subsubsection{Error Analysis}
\begin{figure*}
    \centering
    \includegraphics[scale=0.4]{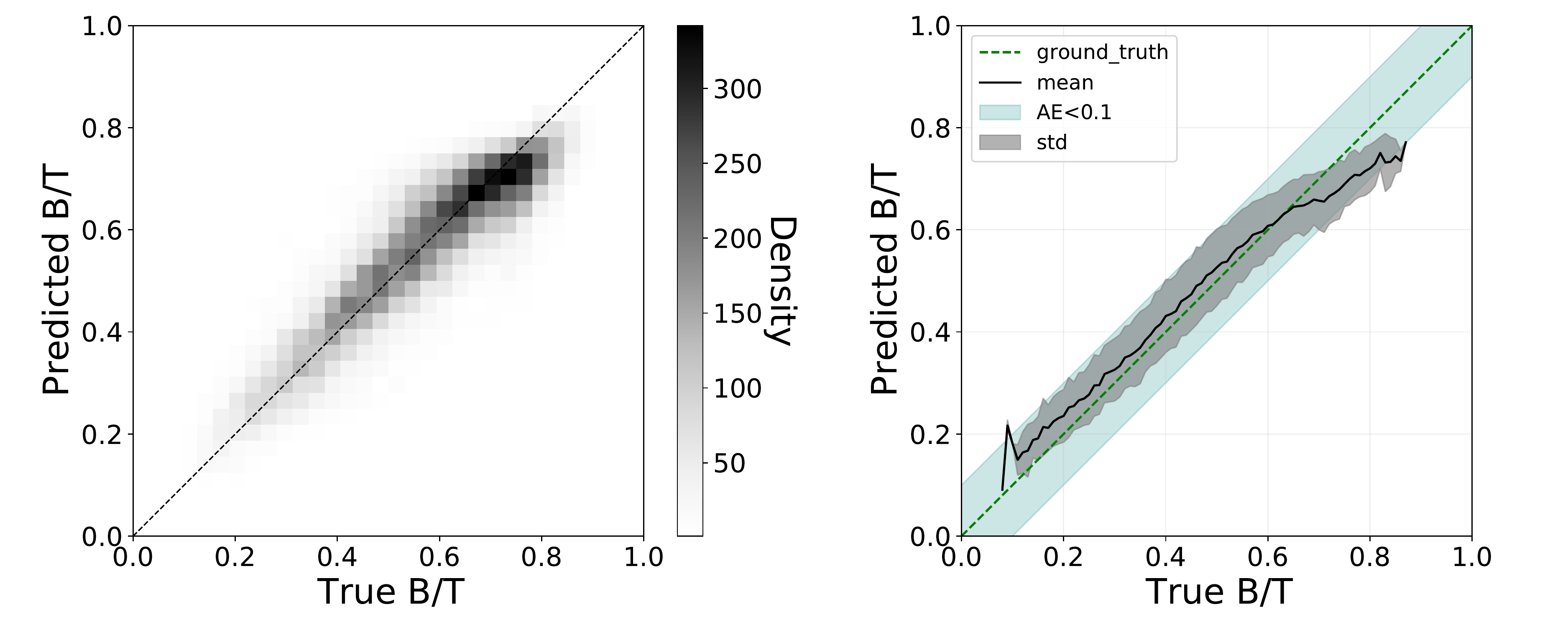}
    \caption[Real values vs Predicted values for 20000 test samples.]{The first plot shows binned 2D histogram of true values vs predicted values for 20000 test samples. Bin size is $0.03\times0.03$. The dashed line shows ground truth, i.e. $AE=0$. We get MAE=0.0543 for the main model. The second plot shows mean and standard deviation for predicted $B/T$ values. The green line shows ground-truth. The light green shaded region shows AE$<$0.1. The black line shows the mean of predicted $B/T$ for each $B/T$ value in test set. The gray shaded region shows standard deviation of predicted $B/T$ for each $B/T$ value in the test set.}
    \label{fig:rvp}
\end{figure*}

The predicted versus true $B/T$ values, as well as the mean and standard deviation of predicted samples are shown in Figure \ref{fig:rvp}. We observe that the model consistently underestimates the $B/T$ ratio for samples with true value higher than 0.65.

\begin{figure*}
    \centering
    \includegraphics[scale=0.7]{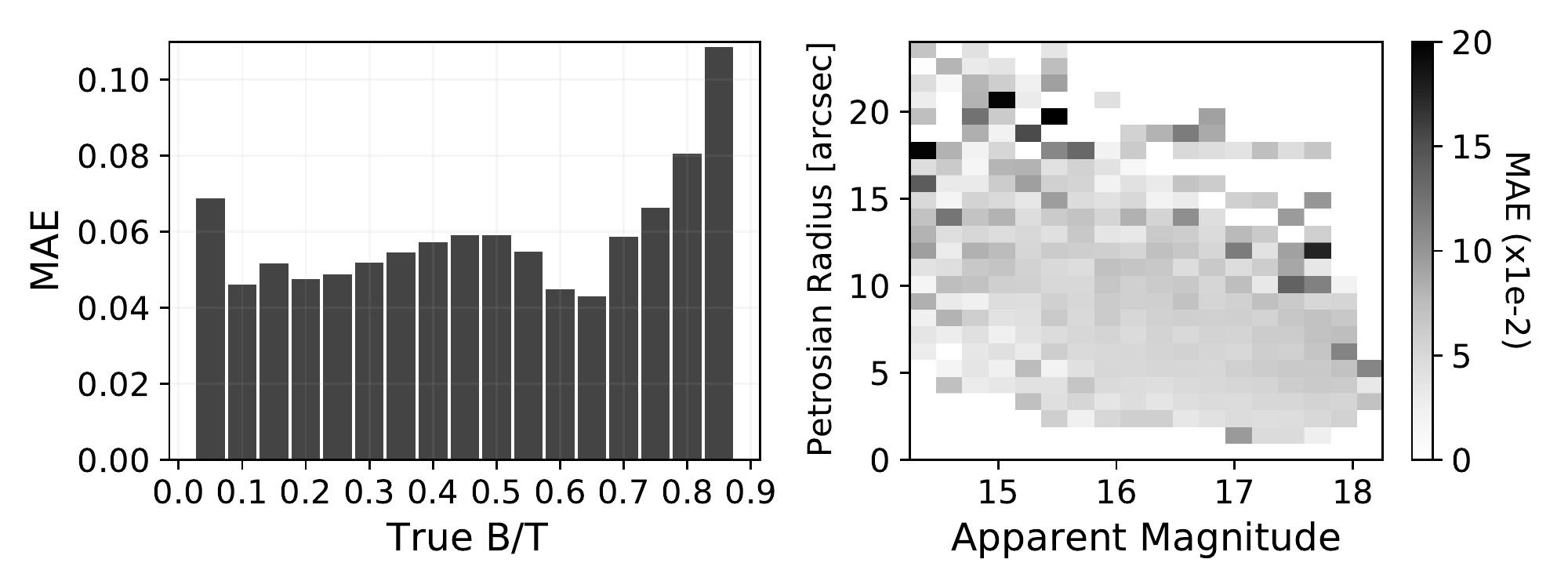}
    \label{fig:mstd(main)}
    \caption[Error plots.]{The first plot shows binned MAE with respect to true $B/T$ values. Bin size used is $0.05$. Higher MAE is observed at higher $B/T$ values. This can be due to lesser training samples for these values. The second plot shows binned MAE with respect to apparent magnitude and petrosian radius of the test samples. Bin size used $0.25\times1$. Again, higher MAE is observed in regions with fewer training samples (see Figure \ref{fig:dist}).}
    \label{fig:err}
\end{figure*}

Error with respect to true $B/T$ can be seen in Figure \ref{fig:err}. It is observed that the number of outliers increases as we move to the edges of the $B/T$ distribution. This is in line with the fact that fewer training samples were present at the edges of the $B/T$ distribution (see Figure \ref{fig:dist}). It is possible that based on our training set there are certain regions in the galaxy size-apparent magnitude relation which are not well trained and hence have higher errors. In Figure \ref{fig:err} we plot the distribution of MAE with galaxy size and apparent magnitude. Notice that the regions of higher error typically lie at the edges which have fewer training samples (see Figure \ref{fig:dist}). Along with these are a few "islands" which are outliers in the size-apparent magnitude relation and thus have quite high errors. The higher signal to noise ratio in the images of brighter galaxies is likely to make the prediction of the $B/T$ ratio with CNN easier. Ignoring galaxies fainter than $r$-band mag of 17 in the test set, we get a sample of 7557 bright galaxies. Accuracy on this set is 87 per cent with MAE=0.0518 using the main model.

\begin{table}
    \centering
    \begin{tabular}{c|c|c|c}
        \hline
        \hline
        Image & True $B/T$ & Pred $B/T$ & AE \\
        \hline
        \hline
        I & 0.79 & 0.73 & 0.06 \\
        \hline
        II & 0.31 & 0.38 & 0.07 \\
        \hline
        III & 0.65 & 0.65 & 0.00 \\
        \hline
        IV & 0.32 & 0.33 & 0.01 \\
        \hline
        V & 0.55 & 0.64 & 0.09 \\
        \hline
        VI & 0.23 & 0.19 & 0.04 \\
        \hline
        VII & 0.58 & 0.60 & 0.02 \\
        \hline
        VIII & 0.25 & 0.36 & 0.11 \\
        \hline
        IX & 0.63 & 0.55 & 0.08 \\
        \hline
        X & 0.28 & 0.30 & 0.02 \\
        \hline
        XI & 0.57 & 0.58 & 0.01 \\
        \hline
        XII & 0.67 & 0.67 & 0.00 \\
        \hline
        XIII & 0.27 & 0.29 & 0.02 \\
        \hline
        XIV & 0.6 & 0.73 & 0.13 \\
        \hline
        XV & 0.56 & 0.54 & 0.04 \\
        \hline
        \hline
    \end{tabular}
    \caption{Details of predictions for samples in Figure \ref{fig:res_main} with the main model. Predicted $B/T$ and AE are rounded off to 2 decimal places.}
    \label{tab:res_main}
\end{table}

\subsubsection{\textbf{Effect of Bright Neighbours}}
With the high sensitivity of modern digital sky surveys, it is quite common to have unrelated stars and galaxies in an image cutout of the galaxy of interest. Occasionally a bright object may lie close to the galaxy of interest. It is important to analyse the effect of bright neighbouring objects on our predictions. To test this, we pick images which have their brightest pixel outside the central region. We define this central region as 15 pixel distance from the centre, or about 6 arcsec. Specifically, if the brightest pixel in an image is more that 6 arcsec away from the centre, we select that image for the bright neighbour set. The 6 arcsec cutoff is based on the median Petrosian radius of our dataset, which is $\sim$4.8 arcsec. The impact of the presence of a bright object on our $B/T$ predictions is shown in Table \ref{tab:bright_neighbours}.
We observe that the performance is reduced by around 3.5 per cent for galaxies with bright neighbouring objects. This also means that our results can be improved further to 86.33 per cent accuracy, if we only consider galaxies with no bright neighbours. Further, ignoring galaxies fainter than $r$-band mag of 17 in the no bright neighbour set, we get a sample of 6460 bright galaxies with no bright neighbours. Accuracy on this set is 87.5 per cent with MAE=0.051. We note that the impact on the accuracy of prediction by considering only bright galaxies and/or those without bright neighbours is fairly small, but measurable.

\begin{table}
    \centering
    \begin{tabular}{c|c|c}
    \hline
    \hline
    & Test samples & Accuracy \\
    \hline
    Overall & 20k & 85.68\% \\
    \hline
    Bright neighbours & $\sim4k$ & 82.92\% \\
    \hline
    No bright neighbours & $\sim16k$ & 86.33\% \\
    \hline
    \hline
    \end{tabular}
    \caption{Accuracy of prediction for  galaxies with and without bright neighbours. The results show that the accuracy is somewhat worse for objects with bright neighbours.}
    \label{tab:bright_neighbours}
\end{table}

\subsubsection{Catastrophic Failures}
We treat galaxies with absolute error in $B/T$ estimation greater than 0.2 as failures. From the 20000 test images, there are a total of 239 failures with the main model. This shows that only $\sim1.2$ per cent of galaxies suffer such failure in our $B/T$ estimation.

\begin{figure*}
    \centering
    \includegraphics[scale=0.75]{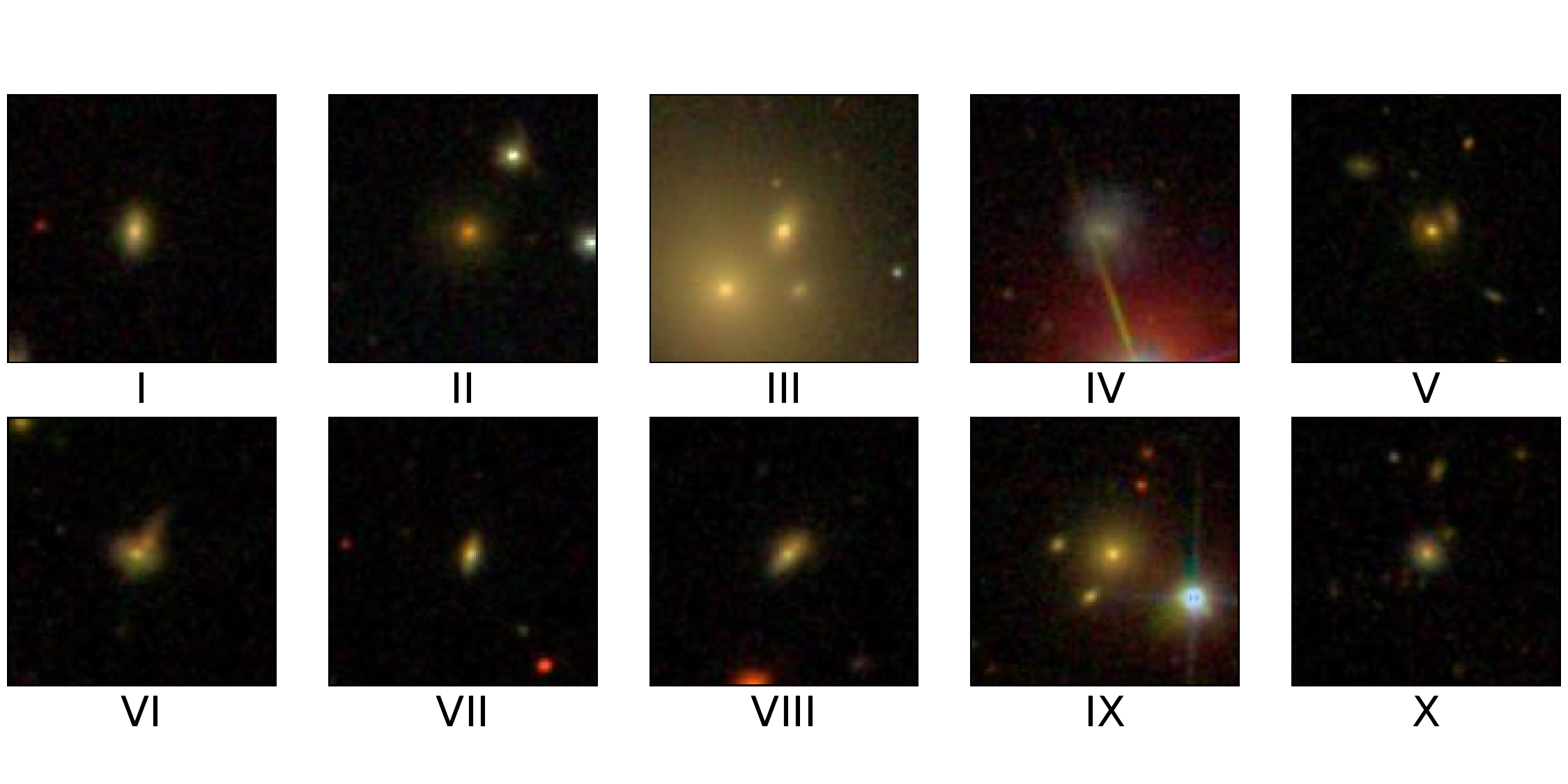}
    \caption{Catastrophic failures with the main model. Refer to Table \ref{tab:cf} for values of True B/T, Predicted $B/T$ and Absolute Error for each image.}
    \label{fig:cf}
\end{figure*}

We treat galaxies with absolute error in $B/T$ estimation greater than 0.3 as catastrophic failures. There are 10 such galaxies using the main model, as shown in Figure \ref{fig:cf} and Table \ref{tab:cf}. This shows that 0.05 per cent of galaxies suffer catastrophic failure in our $B/T$ estimation. We observed no obvious patterns in the properties of the galaxy images which were in the failure or catastrophic failure categories. Many images that were visually similar to the ones present in these failure sets had accurate $B/T$ estimates. Only 3 of these failures had bright neighbours.

\begin{table}
    \centering
    \begin{tabular}{c|c|c|c}
        \hline
        \hline
        Image & True $B/T$ & Pred $B/T$ & AE \\
        \hline
        \hline
        I & 0.81 & 0.49 & 0.32 \\
        \hline
        II & 0.72 & 0.40 & 0.32 \\
        \hline
        III & 0.67 & 0.30 & 0.37 \\
        \hline
        IV & 0.74 & 0.22 & 0.52 \\
        \hline
        V & 0.74 & 0.40 & 0.34 \\
        \hline
        VI & 0.51 & 0.16 & 0.35 \\
        \hline
        VII & 0.76 & 0.45 & 0.31 \\
        \hline
        VIII & 0.64 & 0.34 & 0.30 \\
        \hline
        IX & 0.26 & 0.64 & 0.38 \\
        \hline
        X & 0.73 & 0.32 & 0.41 \\
        \hline
        \hline
    \end{tabular}
    \caption{Catastrophic failures with the main model. Refer to Figure \ref{fig:cf} for the image labels.}
    \label{tab:cf}
\end{table}

\section{Summary and Conclusions} \label{conclusion}

In this work, we have attempted to apply deep learning techniques to a tough regression problem in galaxy morphology. Traditionally, galaxy morphological classification, when done by human experts, is via careful examination of broad-band galaxy images. Inspite of great experience and expertise of the human classifiers, a standard deviation of about 0.5 in the Hubble T-type -- where the same galaxy is classified multiple times by the same expert -- is about the best performance that can be achieved \citep{Nair10}. The alternative to the human expert based approach, is the quantitative bulge-disk decomposition which assumes different analytic profiles for the galaxy components. This approach has the benefit of being quantitative and reproducible, although the time taken for classification scales linearly with the number of galaxies. For human classifiers to guess the $B/T$ luminosity ratio of a galaxy, by visual examination of broad-band images, is considered sufficiently error prone, that it is never attempted in scientific analysis. In this work, we have demonstrated that a task which is considered too difficult for humans, is being performed with reasonably high accuracy by a carefully chosen deep learning network.

We predict the $r$-band $B/T$ ratio of galaxies using end-to-end machine learning, with 85.7 per cent predictions having $AE<0.1$ with no manual intervention. We used the  Xception CNN architecture \citep{xception} with GAP \citep{netinnet} as the machine learning model. We used composite colour images constructed from individual images in the  $(g,r,i)$  bands from SDSS DR15 for training and testing the model. This study only deals with $B/T$ ratio estimates for two-component galaxies as defined in the Meert Catalogue \citep{meert}. The main model takes less than a minute to predict the $B/T$ ratio for 20000 test images. A similar sized sample would take $\sim$100 hours with conventional bulge-disk decomposition pipelines like \textsc{GALFIT} \citep{galfit2002} using wrappers like \textsc{PyMorph} \citep{PYMORPH}, without even including time for data pre-processing. The decomposition codes also require some amount of human intervention. It all adds up to a significant overhead over our method. This overhead would become enormous while predicting properties of billions of galaxies with the next generation of galaxy surveys during the LSST and Euclid eras. Thus our proposed machine learning approach has the potential to save tremendous amount of time, computational and financial resources, along with reduced human effort in predicting an important parameter that characterises galaxies.

\section*{Acknowledgements}

We thank the anonymous referee whose insightful comments helped improve both the content and presentation of this paper. HG thanks NCRA-TIFR for hosting him while the bulk of this work was completed. We would like to thank the Google Colaboratory for providing hardware and software resources for this study. \textsc{Matplotlib} \citep{Hunter2007} was 
used for producing the plots in this paper.

Funding for the Sloan Digital Sky Survey (SDSS) has been provided by the Alfred P. Sloan Foundation, the Participating Institutions, the National Aeronautics and Space Administration, the National Science Foundation, the U.S. Department of Energy, the Japanese Monbukagakusho, and the Max Planck Society. The SDSS Web site is http://www.sdss.org/.

The SDSS is managed by the Astrophysical Research Consortium (ARC) for the Participating Institutions. The Participating Institutions are The University of Chicago, Fermilab, the Institute for Advanced Study, the Japan Participation Group, The Johns Hopkins University, Los Alamos National Laboratory, the Max-Planck-Institute for Astronomy (MPIA), the Max-Planck-Institute for Astrophysics (MPA), New Mexico State University, University of Pittsburgh, Princeton University, the United States Naval Observatory, and the University of Washington.

\section{Data Availability}
To enable readers to obtain $B/T$ ratio for their own sample of galaxies, we are making available the weights and the trained model described in this article. These weights can also be used as galaxy images pre-training for transfer learning. For complete provenance, our image dataset used for training, testing and validation of our model is also being released. These data are available in the following GitHub Repository:  \url{https://github.com/Z3376/Galaxy_BT_Prediction}.





\bibliographystyle{mnras}
\bibliography{bibliography.bib}


\bsp	
\label{lastpage}

\end{document}